# Simulation of Robustness against Lesions of Cortical Networks

**Abbreviated title:** *Simulation of Robustness of Cortical Networks*

Marcus Kaiser[1,2,3,a], Robert Martin[2,4,a], Peter Andras[1,2] and Malcolm P. Young[2]

[1] School of Computer Science, University of Newcastle, Claremont Tower, Newcastle upon Tyne, NE1 7RU, UK

[2] Henry Wellcome Building for Neuroecology, Institute of Neuroscience, University of Newcastle, Framlington Place, Newcastle upon Tyne, NE2 4HH, UK

[3] Jacobs University Bremen, School of Engineering and Science, Campus Ring 6, 28759 Bremen, Germany

[4] FR 2-1, NI, Informatik, Technische Universität Berlin, Franklinstr. 28/29, 10587 Berlin, Germany

[a]Authors contributed equally to this paper

**Correspondence:** Marcus Kaiser; School of Computer Science, University of Newcastle, Claremont Tower, Newcastle upon Tyne, NE1 7RU, UK; E-mail: m.kaiser@ncl.ac.uk

**ABSTRACT**
Structure entails function and thus a structural description of the brain will help to understand its function and may provide insights into many properties of brain systems, from their robustness and recovery from damage, to their dynamics and even their evolution. Advances in the analysis of complex networks provide useful new approaches to understanding structural and functional properties of brain networks. Structural properties of networks recently described allow their characterization as small-world, random (exponential) and scale-free. They complement the set of other properties that have been explored in the context of brain connectivity, such as topology, hodology, clustering, and hierarchical organization. Here we apply new network analysis methods to cortical inter-areal connectivity networks for the cat and macaque brains. We compare these corticocortical fibre networks to benchmark rewired, small-world, scale-free and random networks, using two analysis strategies, in which we measure the effects of the removal of nodes and connections on the structural properties of the cortical networks. The brain networks' structural decay is in most respects similar to that of scale-free networks. The results implicate highly connected hub-nodes and bottleneck connections as structural basis for some of the conditional robustness of brain systems. This informs the understanding of the development of brain networks' connectivity.

**INTRODUCTION**

The brain can be remarkably robust to physical damage. Significant loss of neural tissue can be compensated for in a relatively short time by large-scale adaptation of remaining brain parts (e.g., Spear et al., 1988; Stromswold, 2000; Young, 2000). On the other hand, the removal of small amounts of tissue (e.g. in Broca's area) can lead to a severe functional deficit. These findings provide a somewhat contradictory picture of the robustness of the brain and suggest a number of questions. Can we evaluate effective robustness given this variability in the effects of brain lesions? Are severity and nature of effects of localized damage predictable? We assess here how connectivity data of brain area connectivity can be brought to bear on these questions.

The functionality of any system is grounded in its structural properties. For neurosciences, this has led to exploration of the structural properties of brain networks, such as topology, hodology, clustering, and hierarchical organization (e.g., Nicolelis et al., 1990; Felleman and van Essen, 1991; Young, 1992; Young et al., 1994; Hilgetag et al., 1996; Hilgetag et al., 2000; Sporns et al., 2000; Young, 2000; Petroni et al., 2001; Sporns et al., 2004; Kaiser and Hilgetag, 2006). Recent advances in the study of networks have extended those traditional structural descriptions (Strogatz, 2001), allowing to characterize networks as small-world (Watts and Strogatz, 1998), random and scale-free (e.g., Barabási and Albert, 1999; Albert et al., 2000).

Small-world networks comprise well-connected local neighbourhoods with fewer long-range connections between neighbourhoods. The length of a path between two nodes, that is the number of connections that have to be crossed to go from one node to another, is comparable as low as for a randomly organized network. Scale-free networks are characterized by their specific distribution of connectivities or degrees—the number of connections that each node has. The degree distribution follows a power law. Whereas these networks can have highly-connected nodes or hubs also networks where nodes have maximally 20 connections have been described as scale-free based on the power-law degree distribution (Jeong et al., 2001). Small-world and scale-free properties are compatible, but not equivalent (see, e.g., Amaral et al., 2000).

Scale-free networks have higher robustness than random ones against randomly located damage, whilst being sensitive to damage targeted at their most widely connected nodes (Barabási and Albert, 1999; Young, 2000). This is reminiscent of the properties of the brain described above.

Previous studies have shown that functional networks of the human brain are scale-free (Eguiluz et al., 2005). However, at the level of resting state networks between cortical areas, it was argued that these networks are not scale-free (Achard et al., 2006). Here we analyze what pattern occurs at the level of structural connectivity. In order to establish, whether the brain has properties of scale-free networks, the integrity and robustness to damage of brain networks' structure is compared to that of benchmark random and scale-free networks (an earlier version of this work had been presented as a conference abstract, see Martin et al., 2001).

**MATERIALS AND METHODS**
*Brain structure connectivity data.* We used macaque and cat cortical inter-areal connectivity data (Young, 1993; Young et al., 1994; Scannell et al., 1995) applying the CoCoMac database for the primate data (Stephan et al., 2001; Kötter, 2004). In both species, the data comprised connections among cortical regions of the neocortex.

For the macaque brain, we considered 66 brain structures with 608 connections between them. In the case of the cat brain, we considered 56 structures and 814 connections. We excluded cross-hemispheric connections. The data was represented as the binary connectivity matrix of a graph. Nodes corresponded to the considered brain structures and edges to the reported connections between them. Note that due to the directed nature of brain connections, the connectivity matrix is not necessarily symmetric and the resulting graph has hence directed edges too. The edge density of the macaque brain graph, that is, the number of reported connections divided by the number of all possible connections, is 26.4% (Tab. 1). For the cat brain, the edge density is 14.2%. There are on average 9.2 connections for each structure in the macaque brain and 14.5 connections for the cat brain structures (see supplementary material for the connectivity matrices).

[Table 1 near here]

*Benchmark networks for comparison.* We constructed rewired, scale-free, small-world, and random networks to match the number of nodes and connections of the corresponding two brain networks (Tab. 1). Figure 1 shows examples of small random and scale-free networks to demonstrate differences in their topology. For random networks, the number of connections of a node is close to the average value over all nodes. For scale-free networks, however, nodes with a much higher number of connections can occur; see hub in Fig. 1b.

[Figure 1 near here]

<u>Rewired networks.</u> For rewired networks, each node has the same number of connections as in the original network, however, targets or sources of connections might have changed.Rewired networks were derived from the original cortical networks of the cat and macaque by exchanging connections so that the total number of connections of each node remained the same (the method for randomization is described in Milo et al., 2002). Whereas the degree distribution remains unchained, the cluster architecture is lost during rewiring. Thus, rewired networks allow looking at effects of the degree distribution alone.

<u>Scale-free networks.</u> The algorithm to generate scale-free benchmark networks is based on Barabási and Albert (1999). However, in a modification of their approach we began with an initial graph of six and eight fully connected nodes respectively for the macaque and cat benchmark networks. This was necessary in order to ensure that the clustering coefficient (average percentage of connections between neighbours of a node; see definition below) of the initial graph matched the highest clustering coefficient found in the corresponding brain network. As proposed by Barabási and Albert (1999), further nodes were added one by one to the graph by preferential attachment. At the beginning of this process, the probability that a new node is connected to an existing node *i* is

$$P(i) = \frac{k_i}{\sum_j k_j},$$

where $k_j$ is the number of connections of the node $j$ (Barabási and Albert, 1999). After establishing a connection to node $i^*$, the probabilities are recalculated to reflect the nature of the scale-free networks: if $i$ is connected to $j$, then it is more likely that $i$ is connected to nodes which are already connected to $j$ and it is less likely that $i$ is connected to nodes which are not connected to $j$. The rescaling was undertaken according to

$$P^*(i) = \begin{cases} k_{i^*} \cdot P(i), \text{if } i \text{ and } i^* \text{ are connected} \\ P(i), \text{if } i \text{ and } i^* \text{ are not connected} \end{cases}, \text{ and}$$

$$P(i) = \frac{P^*(i)}{\sum_j P^*(j)}.$$

The probability for the connections in both directions is the same. We confirmed that this modified routine for generating scale-free networks was able to yield a power-law degree distribution (cf. supplementary material).

Small-world networks. Small-world networks were generated by rewiring regular networks as described in the literature (Watts and Strogatz, 1998). The rewiring probability was adjusted so that the resulting networks had similar clustering coefficient than the respective cortical networks (Tab. 1).

Random networks. Whereas all benchmark networks are generated by a random process, we denote Erdös-Renyi random networks (Erdös and Rényi, 1960) as random networks in the remaining manuscript. Random networks were generated by establishing each potential connection between nodes with probability $p$. This probability was the desired connection density, that means, the connection density of the corresponding brain networks, 14.2% of the number of all possible connections for the macaque and 26.4% for the cat. The degree distribution in these random networks followed a binomial probability distribution. For large numbers of nodes this can be approximated by a Poisson distribution and hence the term 'exponential degree distribution' is also used (Bollobas, 1985).

*Graph similarity.* To assess the discrepancy in connectivity between two networks, first their nodes are permuted according to their number of connections. Second, permutated cortical and benchmark matrices are compared by looking what ratio of directed edges in the adjacency matrix that occurred at the same position in both matrices and the total number of directed edges. This percentage is then the graph similarity $S$ between graph $A$ and $B$ given the number of (directed) connections $|E|$:

$$S = \frac{\sum A \wedge B}{|E|},$$

where $\wedge$ is element-by-element multiplication with an element in the resulting matrix non-zero if both elements are non-zero; $\Sigma$ is the sum of all elements in the matrix and thus yields the number of directed edges existing in both matrices, as these are denoted by a value of one in the matrix. Note, that benchmark networks could be more similar than they appear for this measure as not all possible arrangements of nodes were tested. Testing all possibilities ($10^{92}$ for the macaque and $10^{74}$ for the cat) would have been computationally unfeasible.

*Network characterisation.* The clustering coefficient shows the fragmentation of the network. The coefficient is the ratio of the number of existing edges between neighbours of a node *i* and the number of possible edges between all these neighbours. We considered neighbouring nodes of node *i* to be all those nodes that have incoming or outgoing connections between them and node *i*. If a node *i* has $k_i$ neighbours, then the number of all possible in- and outgoing edges between the neighbouring nodes is $k_i * (k_i - 1)$. The coefficient itself is a local property of each node and we define the average coefficient of all nodes to be the clustering coefficient of the graph. This is a measure of how well connected the nodes of the network are.

Following Albert et al. (2000), we considered the average shortest path (ASP) or characteristic path length to characterize the network connectivity and integrity. The ASP between any two nodes in the network is the number of sequential connections that are necessary, on average, to link one node to another by the shortest possible route (Diestel, 1997). In case a network becomes disconnected in the process of removing edges/nodes and there is no path between two nodes, the pair of nodes is ignored. If no two connected points are left, the average shortest path is set to zero. We used Floyd's algorithm to determine the matrix of the shortest paths between each pair of nodes (Cormen et al., 2001). Note that due to directed edges, the shortest path from node *i* to node *j* may not be the same as that from node *j* to node *i*.

*Target determination.* In order to determine the importance of a node to the overall network structure, a simple metric has been used, namely the number of connections formed by this node. In experiments requiring the targeted removal of nodes from the networks, the most highly connected node was eliminated.

To provide the corresponding metric for the targeted elimination of connections (edges) from the network, we chose edge betweenness (Girvan and Newman, 2002), that is, the number of shortest paths between all pairs of nodes that pass though the edge. Edges with high edge betweenness are chosen for targeted attack. Indeed, edge betweenness has been shown to highly correlate with structural network damage for cortical as well as other biological networks (Kaiser and Hilgetag, 2004).

*Analysis methods.* We used the iterative *random* and *targeted* removal of *nodes* and *connections* to analyze the robustness of the networks against damage. Random removal means that we selected a node or connection and deleted it from the graph irrespective of the degree of the node. In the case of targeted removal, we selected the most important node or connection left in the network (see above). After each deletion, we calculated the ASP of the resulting graph. We continued the removal of nodes or connections until all nodes were removed from the network. To derive estimates of the variability in these connectivity measures, we considered 50 benchmark networks for each condition. In the cases of random removal, we repeated the analysis for the brain networks 50 times as well.

# RESULTS
## Degree distribution of cortical networks

Fig. 2 shows the degree distributions of macaque and cat compared with a distribution of random networks. In comparison to random networks, the macaque cortical network has highly connected nodes but also more sparsely connected nodes, reminiscent of scale-free networks. This is also true for the cat network that shows a remarkable number of areas with few connections compared to random networks. Table 2 shows the five most-highly connected nodes for the cat and macaque networks.

[Table 2 near here]

The standard way of observing whether the cortical network resembles a scale-free network would be to search for a power-law in the degree distribution. However, this approach would be inappropriate for cortical networks for three reasons. First, the maximum number of connections of a node equals the number of regions in the network minus one, that means, 65 (macaque) or 55 (cat). Therefore, the degree distribution only consists of two scales. Second, where degree distributions with a low maximal degree had been studied before (Jeong et al., 2001), the number of nodes was considerably higher (>1,800). As less than 100 degrees form the degree distribution, results are unlikely to be robust. Third, there exists a sampling problem in that the amount of unknown or not included connections might change the shape of the degree distribution (Stumpf et al., 2005). Therefore, we will use indirect measures to determine whether cortical networks are similar to scale-free networks.

[Figure 2 near here]

## Graph similarity

Whereas the degree distribution is an abstraction of the underlying network, we looked at a direct comparison between the cortical and benchmark networks. Whereas a direct measure of network similarity was computationally unfeasible (see Methods), we compared the adjacency matrices after ordering nodes by their degrees (see methods). We then looked at the similarity of cortical networks with different benchmark networks (Fig. 3). For rewired cortical networks, the percentage of identical edges was 23% for rewired macaque and 38% for the rewired cat network. Interestingly, benchmark scale-free networks are as similar to the cortical networks as the rewired cortical networks. In contrast, the similarity of random and small-world networks is significantly lower. This can be attributed to the degree distribution of scale-free and cortical networks being comparable as the rewired network only has the degree distribution in common with the original cortical network. After these structural properties, we tested the effect of topological changes on general network properties.

[Figure 3 near here]

## Sequential elimination of nodes

We tested the influence of sequential node elimination on the network structure. Nodes were removed one by one from the network, either randomly or targeted. Plotting the ASP as a function of the fraction of deleted nodes illustrates the characteristic structural disintegration of each network type (See Fig. 4 for the example of targeted elimination of nodes from the Macaque benchmark networks. The complete set of curves for the different analysis types is available as supplementaryt material).

[Figure 4 near here]

Fig. 4A illustrates the effect of random and targeted removal of nodes from the Macaque brain network. Clearly, the specific decline in ASP is different for the two analysis strategies. Whilst the random removal causes only a slow rise in the ASP, targeted removal of highly connected nodes has a much stronger effect on the network structure of the brain network. After a steep rise in ASP the network fragments into smaller components. The remaining shortest paths, that is the paths between nodes within components, are smaller than in the original network. This process leads to a network with pairs of nodes that are connected to each other but not to other nodes of the network. In these cases, the shortest path decreases to a value of one. Finally, also nodes within pairs are removed leading to an ASP of zero.

Fig. 4B–C contrast this specific curve to those observed when removing nodes from the different benchmark networks in a targeted fashion. Whilst the ASP in the random and small-world networks is hardly affected by the targeted elimination of a large proportion of nodes, in the scale-free, like in the brain networks, the effect of targeted node elimination manifests itself in a sharp rise in this measure. Moreover, both, the scale-free and the brain networks show a decline in the ASP around the fraction of deletions, and the characteristic behaviour of the brain network is within the 95% confidence interval encountered for the set of scale-free benchmark networks. This is not the case for the other benchmark networks considered (see Fig. 4).

For the cat brain network (Fig. 5), the random and small-world networks show a different behaviour for targeted node removal than the original cortical network. Though the cat response to targeted node removal is largely within the 95% confidence interval for the scale-free benchmark networks, the peak ASP value and the fraction of deleted nodes where the peak occurs is lower for the cat cortical network.

[Figure 5 near here]

The decline in ASP at a later stage during the elimination process, as observed for the brain and scale-free networks may appear unusual and deserves some additional attention. It can have two reasons. First, it could be that the network gets fragmented into different disconnected components. Each of these is smaller, and likely to have a shorter ASP. Second, the overall decrease in network size with successive eliminations can lead to a decrease in shortest path. This is, however, likely to be a slow process, as it will usually be offset by an increase in ASP due to the targeted nature of the elimination.

In order to quantitatively compare the different graphs, we consider two measures. The first is the maximal ASP measured during the removal of nodes; the second is the fraction of deleted nodes, for which the peak ASP occurs (Fig. 6). For the fraction of peak ASP, only the scale-free benchmark networks are close to the cortical fraction whereas all other benchmark networks show significantly higher fractions. This means that both in the cortical as well as the scale-free networks the removal of highly-connected nodes leads to a rapid increase of ASP so that the fraction of deleted nodes at which the maximum ASP occurs is earlier than for other networks. However, the peak value for scale-free networks is greater than that for cortical networks.

[Figure 6 near here]

**Sequential elimination of connections**
We also tested the similarity of sequential connection elimination. Connections were eliminated one after another either randomly or targeted. Full details of the networks disintegration are shown in the supplementary material. Again we compare the maximal ASP measured during the removal of connections and the fraction of deleted connections, for which the peak ASP occurs (Fig. 7).

[Figure 7 near here]

Only the scale-free benchmark networks yield similar values for both the cat and macaque network whereas other networks yield similar values for just one of the cortical networks.

**DISCUSSION**
We have compared brain inter-area connectivity networks with different types of benchmark networks, including random, scale-free, and small-world networks, and found strong indications that the brain connectivity networks share some of their structural properties with scale-free networks. Besides a formal assessment of the network connectivity (degree distribution and graph similarity, see Figures 3 and 4), the analysis is based on a novel approach, which measures the effect of removal of components of the different networks on their structural integrity. In particular, we compared the effect that the removal of nodes and connections had on the ASP found in the brain connectivity networks and their benchmark counterparts. Note, however, that this analysis is based on cortical connectivity within one hemisphere. Connections between hemispheres and between the cortex and subcortical structures such as thalamic regions were not included. The reason for the lack of interhemispheric connections was that few tracing studies tested for and thus reported fibre tracts towards the contralateral hemisphere. Whereas information about thalamocortical connections would have been available, regions with available information about fibre tracts differed between the cat and macaque. To be consistent between species, the data was not included. For each species, an inclusion of these regions yielded similar results concerning the removal of nodes or edges (supplementary material).

**Simulated robustness and its relation to lesion studies**
How do our simulations relate to experimental lesion studies? Node elimination corresponds indirectly to inactivation or lesion of the corresponding brain areas, and from this perspective, we can interpret this analysis in terms of the brain's robustness to regional damage. The elimination of connections corresponds indirectly to localized brain lesions that damage the white matter and interrupt communication between normally connected brain structures. The ASP yields a measure how well the brain is connected and how well different streams of information can be integrated. Analysing the spatial organisation of cortical networks shows that the brain is optimized towards a low ASP (Kaiser and Hilgetag, 2006). A recent clinical study of the EEG correlation network in Alzheimer patients suggests that increases in ASP lead to a reduced performance in memory tasks (Stam et al., 2007). In this study, the ASP of the EEG synchronization network has been higher in Alzheimer patients compared to the control group. Furthermore, there was a negative correlation between the patients' ASP and their performance in a standard clinical memory test. Whereas the study was based on functional rather structural/anatomical networks, recent studies using diffusion tensor imaging have shown that changes in brain connectivity can be linked to diseases such as Schizophrenia and Alzheimer.

All observations have been made equally during the analysis of the brain networks of cat and macaque, despite different edge densities in the two networks. It is therefore prudent to conclude that it may be extended to other mammalian brain networks. Hence, conditional robustness of brain function may be based to a large extent on two fairly simple structural properties of brain networks: firstly, the number of connections of individual nodes (Young, 2000), i.e., their scale-free nature, and secondly, the heavily connected local clusters with fewer important 'bottlenecks' between them (Kaiser and Hilgetag, 2004). Consequently, it appears feasible to determine the brain structures that are the most important to the maintenance of network function. Typically, brain networks should be able to function robustly in the face of damage to structures that have few connections and damage to connections that do not form part of many shortest connections between pairs of areas. On the other hand, functional effects should be dramatic when structures with very many connections (hubs) are damaged and when connections between structures with very different connectivity patterns (large edge betweenness, cf. Girvan and Newman, 2002) are damaged.

**Is the brain a scale-free network?**
One important feature of our approach is that the rigorous checking of a series of benchmark networks allows assessing the significance of any similarities to other network types found. In the study of a much simpler brain network, it has previously been established that the brain of *C. elegans* is small-world, but not scale-free (Amaral et al., 2000). However, we found that effects of damage on the modelled cat and macaque brain connectivity networks are largely similar to those observed in scale-free networks. Furthermore, the similarity of scale-free and original cortical networks, as measured by graph similarity, was higher than for other benchmark networks. This agrees with other findings: a scale-free network architecture has been found for functional brain networks in humans (Eguiluz et al., 2005). In addition, the human resting state network of 90 cortical and subcortical regions showed similar behaviour after the removal of nodes than our structural network (Achard et al., 2006). This could now be explained by the underlying structural connectivity.

We note that this issue remains controversial. A study of the human resting state network between cortical areas (Achard et al., 2006), concluded that the resting state network is *not* a scale-free network as (a) it is more resilient towards targeted attack compared to a scale-free benchmark network, (b) the degree distribution is not a power-law, and (c) late developing areas such as the dorsolateral prefrontal cortex are among the hubs of the network. The structural network that we analyzed, however, differed from the resting state functional network. First, the resilience towards targeted attack was comparable with that of a scale-free network. Second, though the degree does not follow a power-law distribution this might be due to the small size of the network and incomplete sampling of connections between regions.

**A design for robustness or by-product of functional constraints?**
Is the brain optimized for high robustness or is robustness a by-product of other constraints? In our view, the emergence of highly-connected areas is more likely to be a side effect of brain evolution and development generating structures for efficient processing. For example, highly-connected areas (hubs) in the brain could play a functional role as integrators or spreaders of information (Sporns and Zwi, 2004).

What could be developmental reasons for some regions having a higher connectivity than others? There are several potential developmental mechanisms for yielding brain networks with the highly-connected nodes. Work in brain evolution suggests that when new functional structures are formed by specialization of parts of phylogenetically older structures, the new structures largely inherit the connectivity pattern of the parent structure (e.g., Preuss, 2000). This means that the patterns are repeated and small modifications are added during the evolutionary steps that can arise by duplication of existing areas (Krubitzer and Kahn, 2003). Such inheritance of connectivity by copying of modules is proposed to lead to scale-free metabolic systems (Ravasz et al., 2002). A developmental mechanism for varying the edge degree of regions could be the width of the developmental time window for synaptogenesis at different regions (Kaiser and Hilgetag, 2007).

In conclusion, we have introduced a quantitative method to characterize the robustness of brain networks and compare it to that of standard network types. We have shown that cortical networks are affected in ways similar to scale-free networks concerning the elimination of nodes or connections. However, a direct comparison of degree distributions has been impossible. Our analysis can be extended to employ more elimination strategies or use different properties to characterize the damaged networks. In the future, it would be interesting to compare the effect of experimental lesions with the simulated lesions of our approach. We therefore hope that this theoretical approach will prove useful in modelling robustness towards lesions.

**Acknowledgements**
Supported by the Wellcome Trust, EU Framework Five (R.M) as well as German National Merit Foundation and Fritz-ter-Meer-Foundation (M.K.).

**Abbreviations**
ASP, Average Shortest Path;
ORI, Original (brain) network;
RND, (Erdös-Renyi) Random network;
SF, Scale-free network;
SW, Small-world network;

# References

Achard S, Salvador R, Whitcher B, Suckling J, Bullmore E (2006) A resilient, low-frequency, small-world human brain functional network with highly connected association cortical hubs. J Neurosci 26:63-72.

Albert R, Jeong H, Barabási A-L (2000) Error and Attack Tolerance of Complex Networks. Nature 406:378-382.

Amaral LAN, Scala A, Barthélémy M, Stanley HE (2000) Classes of small-world networks. Proc Natl Acad Sci 97:11149-11152.

Barabási A-L, Albert R (1999) Emergence of Scaling in Random Networks. Science 286:509-512.

Bollobas B (1985) Random Graphs.

Cormen TH, Leiserson CE, Rivest RL, Stein C (2001) Introduction to Algorithms.

Diestel R (1997) Graph Theory. New York: Springer.

Eguiluz VM, Chialvo DR, Cecchi GA, Baliki M, Apkarian AV (2005) Scale-free brain functional networks. Phys Rev Lett 94:018102.

Erdös P, Rényi A (1960) On the evolution of random graphs. Publ Math Inst Hung Acad Sci 5:17-61.

Felleman DJ, van Essen DC (1991) Distributed hierarchical processing in the primate cerebral cortex. Cereb Cortex 1:1-47.

Girvan M, Newman MEJ (2002) Community Structure in Social and Biological Networks. Proc Natl Acad Sci 99:7821-7826.

Hilgetag CC, O'Neill MA, Young MP (1996) Indeterminancy of the visual cortex. Science 271:776-777.

Hilgetag CC, Burns GAPC, O'Neill MA, Scannell JW, Young MP (2000) Anatomical Connectivity Defines the Organization of Clusters of Cortical Areas in the Macaque Monkey and the Cat. Phil Trans R Soc Lond B 355:91-110.

Jeong H, Mason SP, Barabási A-L, Oltvai ZN (2001) Lethality and centrality in protein networks. Nature 411:41-42.

Kaiser M, Hilgetag CC (2004) Edge vulnerability in neural and metabolic networks. Biol Cybern 90:311-317.

Kaiser M, Hilgetag CC (2006) Nonoptimal Component Placement, but Short Processing Paths, due to Long-Distance Projections in Neural Systems. PLoS Computational Biology 2:e95.

Kaiser M, Hilgetag CC (2007) Development of multi-cluster cortical networks by time windows for spatial growth. Neurocomputing:(in press).

Kötter R (2004) Online Retrieval, Processing, and Visualization of Primate Connectivity Data from the CoCoMac Database. Neuroinformatics 2:127-144.

Krubitzer L, Kahn DM (2003) Nature versus Nurture Revisited: An Old Idea with a New Twist. Prog Neurobiol 70:33-52.

Martin R, Kaiser M, Andras P, Young MP (2001) Is the Brain a Scale-Free Network? In: Annual Conference of the Society for Neuroscience, p Paper 816.814. San Diego, US.

Milo R, Shen-Orr S, Itzkovitz S, Kashtan N, Chklovskii D, Alon U (2002) Network Motifs: Simple Building Blocks of Complex Networks. Science 298:824-827.

Nicolelis MAL, Yu CH, Baccalá LA (1990) Structural Characterization of the Neural Circuit responsible for Control of the Cardiovascular Functions in High Vertebrates. Comput Biol Med 20:379-400.

Petroni F, Panzeri S, Hilgetag CC, Koetter R, Young MP (2001) Simultaneity of Responses in a Hierarchical Visual Network. Neuroreport 12:2753-2759.

Preuss TM (2000) What's human about the human brain. In: The New Cognitive Neurosciences (Gazzaniga M, ed), pp 1219-1234. Cambridge, MA.

Ravasz E, Somera AL, Mongru DA, Oltvai ZN, Barabási A-L (2002) Hierarchical Organization of Modularity in Metabolic Networks. Science 297:1551-1555.

Scannell JW, Blakemore C, Young MP (1995) Analysis of Connectivity in the Cat Cerebral Cortex. J Neurosci 15:1463-1483.

Spear PD, Tong L, McCall MA (1988) Functional influence of areas 17, 18 and 19 on lateral suprasylvian cortex in kittens and adult cats: implications for compensation following early visual cortex damage. Brain Res 447:79-91.

Sporns O, Zwi JD (2004) The Small World of the Cerebral Cortex. Neuroinformatics 2:145-162.

Sporns O, Tononi G, Edelman GM (2000) Theoretical Neuroanatomy: Relating Anatomical and Functional Connectivity in Graphs and Cortical Connection Matrices. Cereb Cortex 10:127-141.

Sporns O, Chialvo DR, Kaiser M, Hilgetag CC (2004) Organization, development and function of complex brain networks. Trends Cogn Sci 8:418-425.

Stam CJ, Jones BF, Nolte G, Breakspear M, Scheltens P (2007) Small-world networks and functional connectivity in Alzheimer's disease. Cereb Cortex 17:92-99.

Stephan KE, Kamper L, Bozkurt A, Burns GA, Young MP, Kotter R (2001) Advanced database methodology for the Collation of Connectivity data on the Macaque brain (CoCoMac). Philos Trans R Soc Lond B Biol Sci 356:1159-1186.

Strogatz SH (2001) Exploring complex networks. Nature 410:268-276.

Stromswold K (2000) The cognitive neuroscience of language acquisition. In: The New Cognitive Neurosciences (Gazzaniga M, ed), pp 909-932. Cambridge, MA.

Stumpf MPH, Wiuf C, May RM (2005) Subnets of Scale-Free Networks are Not Scale-Free: Sampling Properties of Networks. Proc Natl Acad Sci USA 102:4221-4224.

Watts DJ, Strogatz SH (1998) Collective Dynamics of 'small-World' Networks. Nature 393:440-442.

Young MP (1992) Objective Analysis of the Topological Organization of the Primate Cortical Visual System. Nature 358:152-155.

Young MP (1993) The organization of neural systems in the primate cerebral cortex. Phil Trans R Soc 252:13-18.

Young MP (2000) The architecture of visual cortex and inferential processes in vision. Spatial Vision 13:137-146.

Young MP, Scannell JW, Burns GA, Blakemore C (1994) Analysis of connectivity: neural systems in the cerebral cortex. Rev Neurosci 5:227-250.

# Tables

**Table 1. Comparison of brain networks and benchmark networks.**
The table shows the average shortest path and the clustering coefficient statistics for the macaque and cat brain structure networks, and for the respective benchmark random, rewired, small-world, and scale-free networks. For the benchmark networks, the data shows the mean value and the standard deviation of 50 generated networks.

|  | Average shortest path | Clustering coefficient |
|---|---|---|
| **Macaque** | 2.414 | 0.453 |
| Random mean | $2.093 \pm 0.009$ | $0.142 \pm 0.004$ |
| Rewired mean | $2.118 \pm 0.010$ | $0.239 \pm 0.009$ |
| Small-world mean | $2.439 \pm 0.054$ | $0.416 \pm 0.022$ |
| Scale-free mean | $2.078 \pm 0.042$ | $0.564 \pm 0.042$ |
| **Cat** | 1.961 | 0.542 |
| Random mean | $1.749 \pm 0.002$ | $0.265 \pm 0.003$ |
| Rewired mean | $1.803 \pm 0.006$ | $0.381 \pm 0.006$ |
| Small-world mean | $1.868 \pm 0.017$ | $0.461 \pm 0.016$ |
| Scale-free mean | $1.768 \pm 0.014$ | $0.535 \pm 0.029$ |

**Table 2. Overview of the most highly-connected regions in the cat and macaque network.**
The table shows the total number of connections of the region (degree) as well as the number of incoming / afferent (in-degree) and outgoing / efferent (out-degree) connections. The maximal possible number of connections would have been 110 connections for the cat and 130 connections for the macaque.

**Cat**

| Rank | Area | Total | Incoming | Outgoing |
|---|---|---|---|---|
| 1 | AES | 59 | 30 | 29 |
| 2 | Ia | 55 | 29 | 26 |
| 3 | 7 | 54 | 28 | 26 |
| 4 | Ig | 52 | 22 | 30 |
| 5 | 5al | 49 | 30 | 19 |

**Macaque**

| Rank | Area | Total | Incoming | Outgoing |
|---|---|---|---|---|
| 1 | A7B | 43 | 23 | 20 |
| 2 | LIP | 42 | 19 | 23 |
| 3 | A46 | 42 | 23 | 19 |
| 4 | FEF | 38 | 19 | 19 |
| 5 | TPT | 37 | 18 | 19 |

**Figures**

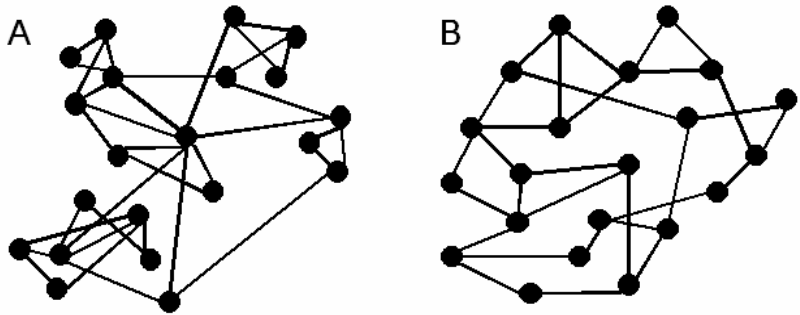

**Figure 1. Examples of random and scale-free networks.** Schematic view of network connectivity features. (A) Simple scale-free network having highly-connected nodes (hubs) here shown at the centre. (B) Simple random network; both networks have the same number of nodes and edges.

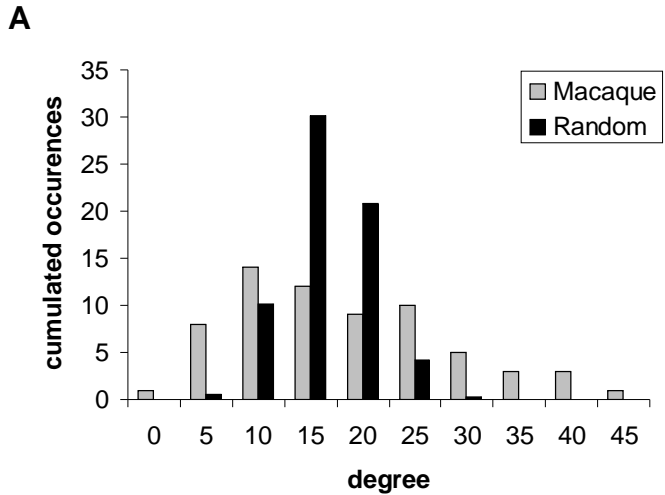

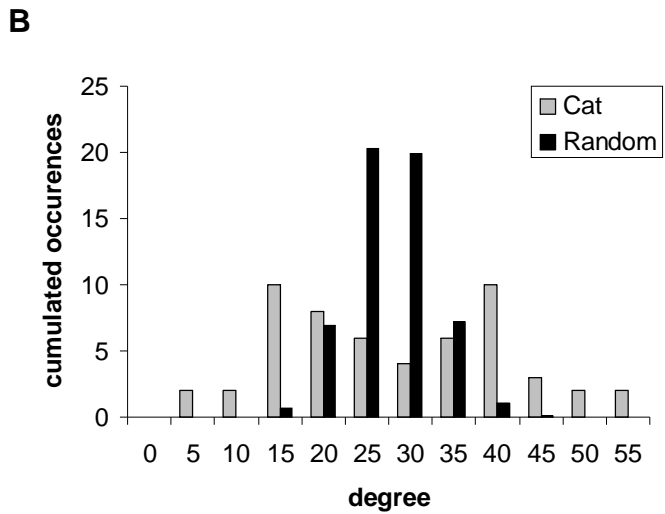

**Figure 2. Direct comparison of degree distribution.** (A) Histogram of the degree distribution of the macaque (gray) compared to the distribution of random networks (binomial distribution given the probability p=0.1417 that an edge occurs, black). (B) Histogram of the degree distribution of the cat (gray) compared to the distribution of random networks (binomial distribution given the probability p=0.2643 that an edge occurs, black).

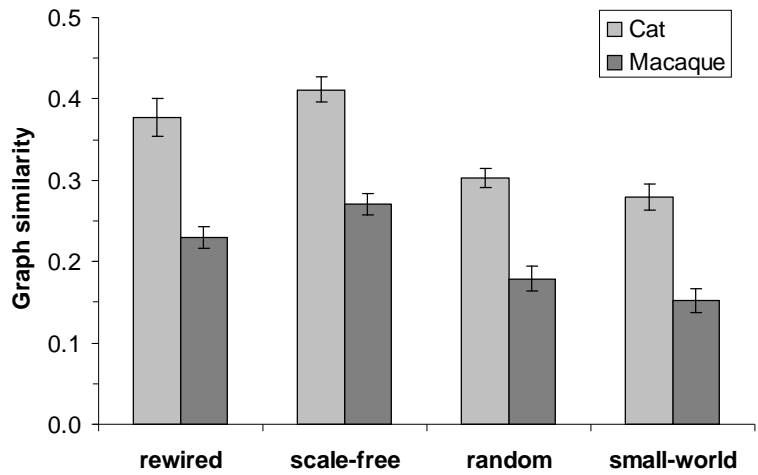

**Figure 3. Similarity of network connectivity.** For each type of benchmark network, 1,000 networks were generated. As the cat network has a larger number of edges, the percentages of similar edges are also higher. The similarity with the cortical networks is as good for the scale-free networks as for the rewired cortical networks. In contrast, the similarity of random and small-world networks is significantly lower.

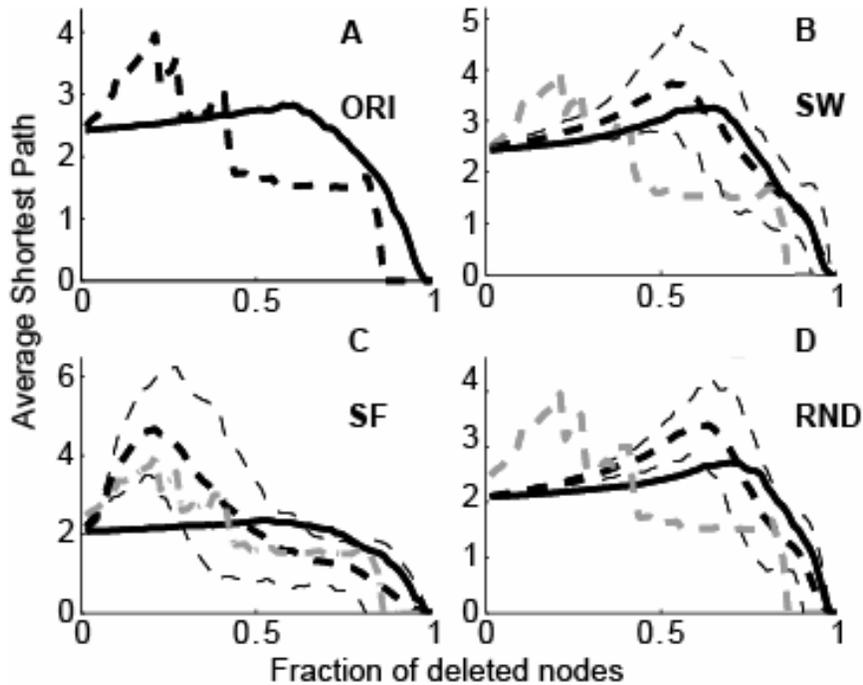

**Figure 4. Sequential node eliminations in Macaque cortical networks.** The fraction of deleted nodes (zero for the intact network) is plotted against the average shortest path (ASP) after node removals. Nodes were removed in order of connectivity, starting with the most highly connected nodes (targeted elimination) or the node order was determined randomly (random elimination). (A) Cortical network during targeted (dashed) and random (solid line) elimination. In the subsequent plots B, C and D, the dashed line shows the average effect of targeted elimination and the thin dashed lines the 95% confidence interval for the generated networks. The solid line represents the average effect of random elimination. The dashed grey line represents targeted removal in the cortical network of A for comparison. (B) Small-world benchmark network. (C) Scale-free benchmark network. (D) Random benchmark network. (The complete set of figures for cat and macaque with node and edge elimination and the effect on ASP is available in the supplementary material).

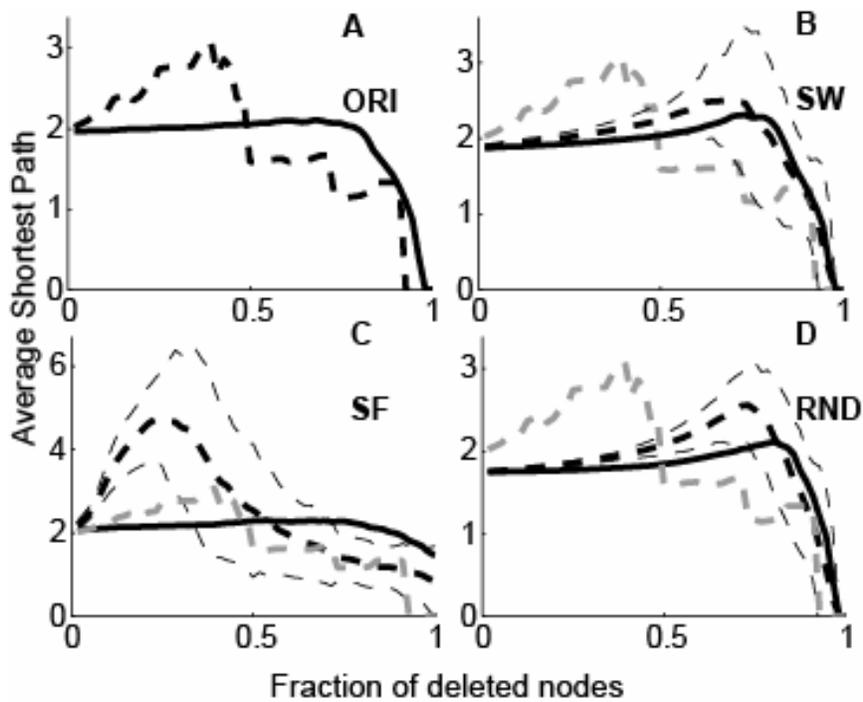

**Figure 5. Sequential node eliminations in cat cortical networks.** The fraction of deleted nodes (zero for the intact network) is plotted against the average shortest path (ASP) after node removals. Nodes were removed in order of connectivity, starting with the most highly connected nodes (targeted elimination) or the node order was determined randomly (random elimination). (A) Cortical network during targeted (dashed) and random (solid line) elimination. Lines in B-C have the same meaning as in Fig. 4. (B) Small-world benchmark network. (C) Scale-free benchmark network. (D) Random benchmark network.

A

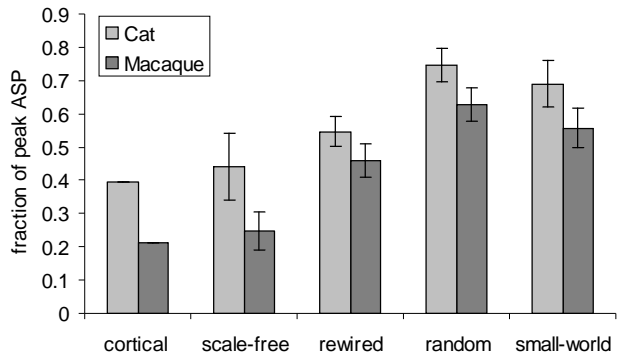

B

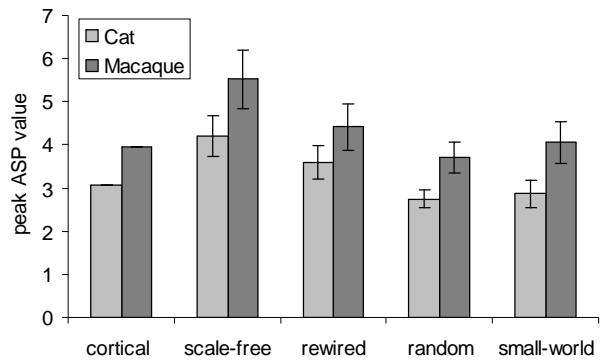

**Figure 6. Fraction and value of peak ASP for targeted node elimination.** The average values and standard deviations are shown for the 50 generated benchmark networks. (A) Fraction of eliminated nodes, at which the largest ASP was attained. For the cat cortical network, only the fraction of peak ASP for the scale-free network is close to the cat network whereas the fractions of other benchmark networks are higher. The same is the case for the macaque cortical network. (B) Peak value of the ASP. It is higher for scale-free networks than for cortical networks, in contrast to more similar values for the other benchmark networks.

**A**

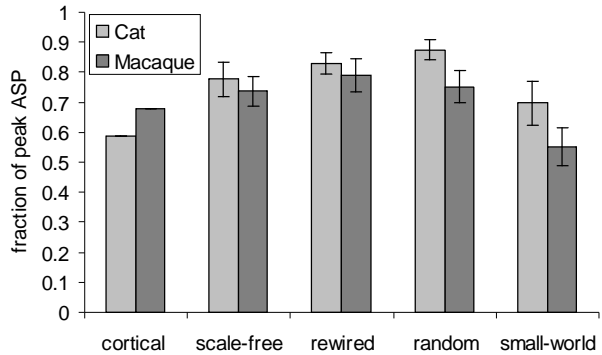

**B**

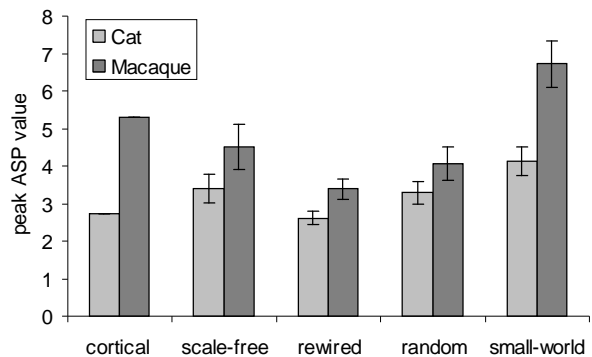

**Figure 7. Fraction and value of peak ASP for targeted connection elimination.** The average values and standard deviations are shown for the 50 generated benchmark networks. (A) Fraction of eliminated connections, at which the largest ASP was attained. For the cat network, scale-free and small-world fractions are similar to the cortical value whereas fractions of rewired and random networks are significantly higher. For the macaque network, however, all benchmark networks except for the small-world network show a similar fraction of peak ASP. (B) Peak values of the ASP. The peak value of the cat cortical network can be matched by the random and rewired networks, nearly by the scale-free but significantly not by the small-world network. For the macaque, all networks except for the scale-free network show significantly different values.